\documentclass[runningheads]{llncs}
\usepackage{graphicx}
\usepackage{amsmath,amssymb} % define this before the line numbering.
\usepackage{wrapfig}
\usepackage{makecell}
\usepackage{color}
\usepackage[width=122mm,left=12mm,paperwidth=146mm,height=193mm,top=12mm,paperheight=217mm]{geometry}
\usepackage{floatrow}
\usepackage{graphicx}
\begin{document}
% \linenumbers
\pagestyle{headings}
\mainmatter
\def\ECCV18SubNumber{}  % Insert your submission number here

\title{HairNet: Single-View Hair Reconstruction using Convolutional Neural Networks} % Replace with your title

\titlerunning{}
\authorrunning{}

\author{Yi Zhou$^1$ \and Liwen Hu$^1$ \and Jun Xing$^2$ \and Weikai Chen$^2$ \and Han-Wei Kung$^{3}$ \and Xin Tong$^4$ \and Hao Li$^{1,2,3}$}
%\institute{University of Southern California\and USC Institute for Creative Technologies \and University of California Santa Barbara \and Pinscreen \and Microsoft}
\institute{$^1$ University of Southern California    $^2$ USC Institute for Creative Technologies\\
$^3$ Pinscreen  $^4$ Microsoft Research Asia}

\maketitle
\vspace{-20px}
\begin{abstract}
We introduce a deep learning-based method to generate full 3D hair geometry from an unconstrained image. Our method can recover local strand details and has real-time performance.
State-of-the-art hair modeling techniques rely on large hairstyle collections for nearest neighbor retrieval and then perform ad-hoc refinement. Our deep learning approach, in contrast, is highly efficient in storage and can run 1000 times faster while generating hair with 30K strands. The convolutional neural network takes the 2D orientation field of a hair image as input and generates strand features that are evenly distributed on the parameterized 2D scalp. 
We introduce a collision loss to synthesize more plausible hairstyles, and the visibility of each strand is also used as a weight term to improve the reconstruction accuracy.
The encoder-decoder architecture of our network naturally provides a compact and continuous representation for hairstyles, which allows us to interpolate naturally between hairstyles.
We use a large set of rendered synthetic hair models to train our network. 
Our method scales to real images because an intermediate 2D orientation field, automatically calculated from the real image, factors out the difference between synthetic and real hairs.
We demonstrate the effectiveness and robustness of our method on a wide range of challenging real Internet pictures, and show reconstructed hair sequences from videos.

\end{abstract}
\begin{figure}[h!]
    \centering
    \includegraphics[width=1.0\textwidth]{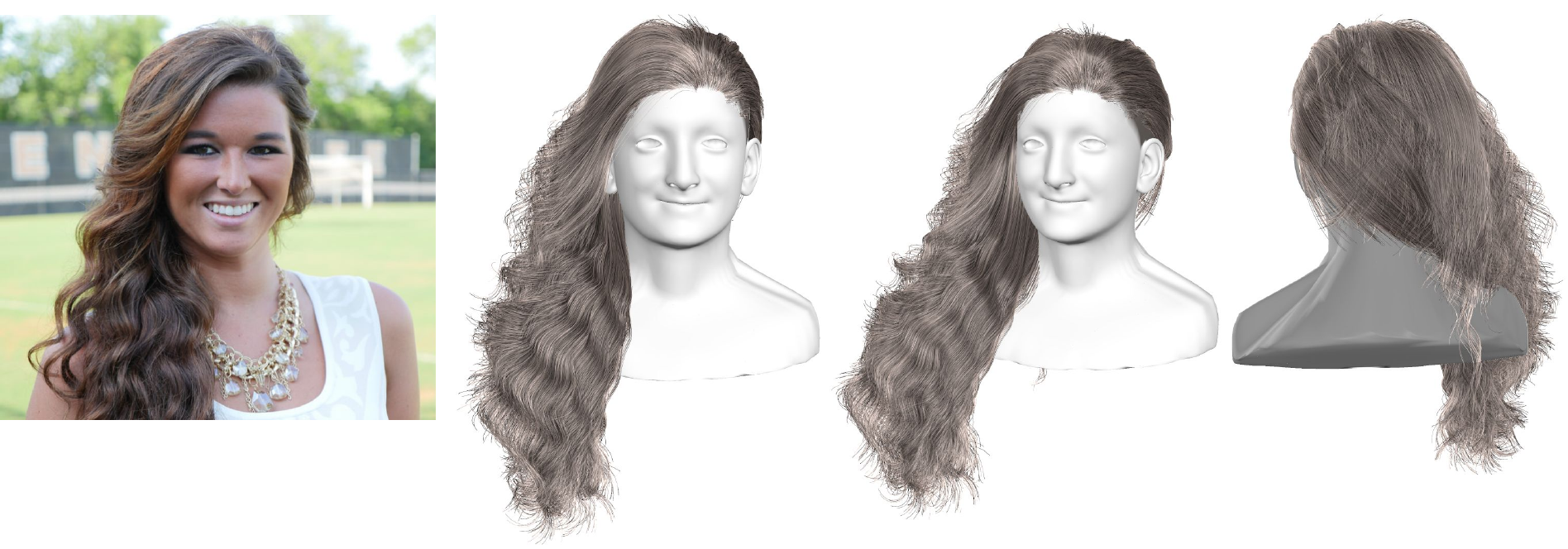}
    \vspace{-10px}
    \caption{Hair reconstruction from a single view image using HairNet.}
    \label{fig:teaser}
\end{figure}

\section{Introduction}

Realistic hair modeling is one of the most difficult tasks when digitizing virtual humans \cite{cao2014face,hu2017avatar,li2015facial,olszewski2016high,hadap2007strands}. In contrast to objects that are easily parameterizable, like the human face, hair spans a wide range of shape variations and can be highly complex due to its volumetric structure and level of deformability in each strand.
Although \cite{Paris:2008:HPG,Jakob:2009:CHA,Bee12,luo2013structure,Xu:2014:DHC} can create high-quality 3D hair models, but they require specialized hardware setups that are difficult to be deployed and populated.
Chai et al.~\cite{Chai:2013:DHM,Chai:2012:SHM} introduced the first simple hair modeling technique from a single image, but the process requires manual input and cannot properly generate non-visible parts of the hair. Hu et al. \cite{hu2015single} later addressed this problem by introducing a data-driven approach, but some user strokes were still required. More recently, Chai et al. \cite{chai2016autohair} adopted a convolutional neural network to segment the hair in the input image to fully automate the modeling process, and \cite{Zhang:2017:DAF} proposed a four-view approach for more flexible control.

%However, these data-driven techniques rely on storing and querying a huge hair model dataset and computationally-heavy search and refinement steps.

However, these data-driven techniques rely on storing and querying a huge hair model dataset and performing computationally-heavy refinement steps.
Thus, they are not feasible for applications that require real-time performance or have limited hard disk and memory space.
More importantly, these methods reconstruct the target hairstyle by fitting the retrieved hair models to the input image, which may capture the main hair shape well, but cannot handle the details nor achieve high accuracy.
Moreover, since both the query and refinement of hair models are based on an undirected 2D orientation match, where a horizontal orientation tensor can either direct to the right or the left, this method may sometimes produce hair with incorrect growing direction or parting lines and weird deformations in the z-axis.

To speed up the procedure and reconstruct hairs that preserve better style w.r.t the input image and look more natural, we propose a deep learning based approach to generate the full hair geometry from a single-view image, as shown in Figure \ref{fig:teaser}.
Different from recent advances that synthesize shapes in the form of volumetric grids \cite{ChoyXGCS16} or point clouds \cite{FanSG16} via neural networks, our method generates the hair strands directly, which are more suitable for non-manifold structures like hair and could achieve much higher details and precision.

Our neural network, which we call HairNet, is composed of a convolutional encoder that extracts the high-level hair-feature vector from the 2D orientation field of a hair image, and a deconvolutional decoder that generates $32\times32$ strand-features evenly distributed on the parameterized 2D scalp.
The hair strand-features could be interpolated on the scalp space to get higher (30K) resolution and further decoded to the final strands, represented as sequences of 3D points.  
In particular, the hair-feature vector can be seen as a compact and continuous representation of the hair model, which enables us to sample or interpolate more plausible hairstyles efficiently in the latent space.
In addition to the reconstruction loss, we also introduce a collision loss between the hair strands and a body model to push the generated hairstyles towards a more plausible space.
To further improve the accuracy, we uses the visibility of each strand based on the input image as a weight to modulate its loss.

Obtaining a training set with real hair images and ground-truth 3D hair geometries is challenging. We can factor out the difference between synthetic and real hair data by using an intermediate 2D orientation field as network input. This enables our network to be trained with largely accessible synthetic hair models and also real images without any changes.
For example, the 2D orientation field can be calculated from a real image by applying a Gabor filter on the hair region automatically segmented using the method of \cite{zhao2017pspnet}.
Specifically, we synthesized a hair data set composed of 40K different hairstyles and 160K corresponding 2D orientation images rendered from random views for training.

Compared to previous data-driven methods that could take minutes and terabytes of disk storage for a single reconstruction, our method only takes less than 1 second and 70 MB disk storage in total.
We demonstrate the effectiveness and robustness of our method on both synthetic hair images and real images from the Internet, and show applications in hair interpolation and video tracking.

Our contributions can be summarized as follows:
\begin{enumerate}
\item 
We propose the first deep neural network to generate dense hair geometry from a single-view image. 
To the best of our knowledge, it is also the first work to incorporate both collision and visibility in a deep neural network to deal with 3D geometries.
\item
Our approach achieves state-of-the-art resolution and quality, and significantly outperforms existing data-driven methods in both speed and storage.
\item
Our network provides the first compact and continuous representation of hair geometry, from which different hairstyles can be smoothly sampled and interpolated.
\item
We construct a large-scale database of around 40K 3D hair models and 160K corresponding rendered images.
\end{enumerate}

\section{Related Work}

\paragraph{Hair Digitization.}
A general survey of existing hair modeling techniques can be found in Ward et.al \cite{Ward06asurvey}.
For experienced artists, purely manual editing from scratch with commercial softwares such as XGen and Hairfarm is chosen for highest quality, flexibility and controllability, but the modeling of compelling and realistic
hairstyles can easily take several weeks.
To avoid tedious manipulations on individual hair fibers, some efficient design tools are proposed in \cite{Choe:2005:ASW,Kim:2002:IMH,Fu:2007:SH,Yuksel:2009:HM,Weng:2013:HIP}.

Meanwhile, hair capturing methods have been introduced to acquire hairstyle data from the real world. Most hair capturing methods typically rely on high-fidelity acquisition systems, controlled recording sessions, manual assistance such as multi-view stereo cameras\cite{Paris:2008:HPG,Bee12,Jakob:2009:CHA,luo2013structure,Echevarria:2014:CSH,Xu:2014:DHC,hu2014robust}, single RGB-D camera \cite{hu2014capturing} or thermal imaging \cite{Herrera:2012:LHI}.

More recently, Single-view hair digitization methods have been proposed by Chai et.al \cite{Chai:2012:SHM,Chai:2013:DHM} but can only roughly produce the frontal geometry of the hair. 
Hu et.al \cite{hu2015single} later demonstrated the first system that can model entire hairstyles at the strand level using a database-driven reconstruction technique with minimal user interactions from a single input image. A follow-up automatic method has been later proposed by \cite{chai2016autohair}, which uses a deep neural network for hair segmentation and augments a larger database for shape retrieval.
To allow more flexible control of side and back views of the hairstyle, Zhang et.al \cite{Zhang:2017:DAF} proposed a four-view image-based hair modeling method to fill the gap between multi-view and single-view hair capturing techniques.
Since these methods rely on a large dataset for matching, speed is an issue and the final results depend highly on the database quality and diversity.

\paragraph{Single-View Reconstruction using Deep Learning.}

Generation of 3D data by deep neural networks has been attracting increasing attention recently.
Volumetric CNNs \cite{ChoyXGCS16,GirdharFRG16,TulsianiZEM17,JacksonBAT17} use 3D convolutional neural networks to generate voxelized shapes but are highly constrained by the volume resolution and computation cost of 3D convolution.
Although techniques such as hierarchical reconstruction \cite{HaneTM17} and octree \cite{riegler2017octnet,tatarchenko2017octree,Wang:2017:OOC} could be used to improve the resolution, generating details like hair strands are still extremely challenging.

On the other hand, point clouds scale well to high resolution due to their unstructured representation.
\cite{qi2016pointnet,qi2017pointnet++} proposed unified frameworks to learn features from point clouds for tasks like 3D object classification and segmentation, but not generation. 
Following the pioneering work of PointNet, \cite{guerrero2017pcpnet} proposed the PCPNet to estimate the local normal and curvature from point sets, and \cite{FanSG16} proposed a network for point set generation from a single image.
However, point clouds still exhibit coarse structure and are not able to capture the topological structure of hair strands.

\section{Method}

The entire pipeline contains three steps.
A preprocessing step is first adopted to calculate the 2D orientation field of the hair region based on the automatically estimated hair mask. Then, HairNet takes the 2D orientation fields as input and generates hair strands represented as sequences of 3D points. A reconstruction step is finally performed to efficiently generate a smooth and dense hair model.
\vspace{-6px}

\subsection{Preprocessing}
We first adopt PSPNet \cite{zhao2017pspnet} to produce an accurate and robust pixel-wise hair mask of the input portrait image, followed by computing the undirected 2D orientation for each pixel of the hair region using a Gabor filter \cite{luo2013structure}.
The use of undirected orientation eliminates the need of estimating the hair growth direction, which otherwise requires extra manual labeling \cite{hu2015single} or learning \cite{chai2016autohair}.
However, the hair alone could be ambiguous due to the lack of camera view information and its scale and position with respect to the human body. Thus we also add the segmentation mask of the human head and body on the input image.
In particular, the human head is obtained by fitting a 3D morphable head model to the face \cite{hu2017avatar} and the body could be positioned accordingly via rigid transformation.
All these processes could be automated and run in real-time. The final output is a $3\times256\times256$ image, whose first two channels store the color-coded hair orientation and third channel indicates the segmentation of hair, body and background.

\vspace{-6px}
\subsection{Data Generation}
Similar to Hu et. al \cite{hu2015single}, we first collect an original hair dataset with 340 3D hair models from public online repositories \cite{EA2017}, align them to the same reference head, convert the mesh into hair strands and solve the collision between the hair and the body. We then populate the original hair set via mirroring and pair-wise blending. 

Different from AutoHair \cite{chai2016autohair} which simply uses volume boundaries to avoid unnatural combinations, we separate the hairs into 12 classes based on styles shown in table \ref{hairclass} and blend each pair of hairstyles within the same class to generate more natural examples.
In particular, we cluster the strands of each hair into five central strands, and each pair of hairstyles can generate $2^5-2$ additional combinations of central strands. The new central strands serve as a guidance to blend the detailed hairs. 
Instead of using all of the combinations, we randomly select the combination of them for each hair pair, leading to a total number over 40K hairs for our synthetic hair dataset.

\vspace{-10px}
\setlength{\tabcolsep}{0.8em}
\begin{table}[]
\centering
\caption{Hair classes and the number of hairs in each class. S refers to short, M refers to medium, L refers to long, X refers to very, s refers to straight and c refers to curly. Some hairs are assigned to multiple classes if its style is ambiguous.}
\label{hairclass}
\begin{tabular}{|l|l|l|l|l|l|l|l|l|l|l|l|}
\hline
$XS_s$ & 20 & $S_s$ & 110 & $M_s$ & 28 & $L_s$ & 29 & $XL_s$ & 27 & $XXL_s$ & 4 \\ \hline
$XS_c$ & 0  & $S_c$ & 19  & $M_c$ & 65 & $L_c$ & 27 & $XL_c$ & 23 & $XXL_c$ & 1 \\ \hline
\end{tabular}
\end{table}

\vspace{-10px}

In order to get the corresponding orientation images of each hair model, we randomly rotate and translate hair inside the view port of a fixed camera and render 4 orientation images at different views. The rotation ranges from -90$^\circ$ to +90$^\circ$ for the yaw axis and -15$^\circ$ to +15$^\circ$ for the pitch and roll axis. 
We also add Gaussian noises to the orientation to emulate the real conditions.

\vspace{-6px}
\subsection{Hair Prediction Network}

\newcommand{\sample}{s} 
\newcommand{\pos}{\mathbf{p}}
\newcommand{\curv}{c} 

\subsubsection{Hair Representation.}
We represent each strand as an ordered 3D point set $\zeta =\{\sample_{i}\}^{M}_{i=0}$, evenly sampled with a fixed number ($M$ = 100 in our experiments) of points from the root to end.
Each sample $\sample_{i}$ contains attributes of position $\pos_{i}$ and curvature $\curv_{i}$.
Although the strands have large variance in length, curliness, and shape, they all grow from fixed roots to flexible ends.
To remove the variance caused by root positions, we represent each strand in the local coordinate anchored at its root.

The hair model can be treated as a set of $N$ strands $H=\zeta ^N$ with fixed roots, and can be formulated as a matrix $A_{N*M}$, where each entry $A_{i,j} = (\pos_{i,j}, \curv_{i,j})$ represents the $j$th sample point on the $i$th strand. 
In particular, we adopt the method in \cite{Wang:2009:EHG} to parameterize the scalp to a $32\times32$ grid, and sample hair roots at those grid centers ($N$ = 1024).

\begin{figure}[h]
    \centering
    \includegraphics[width=1.0\textwidth]{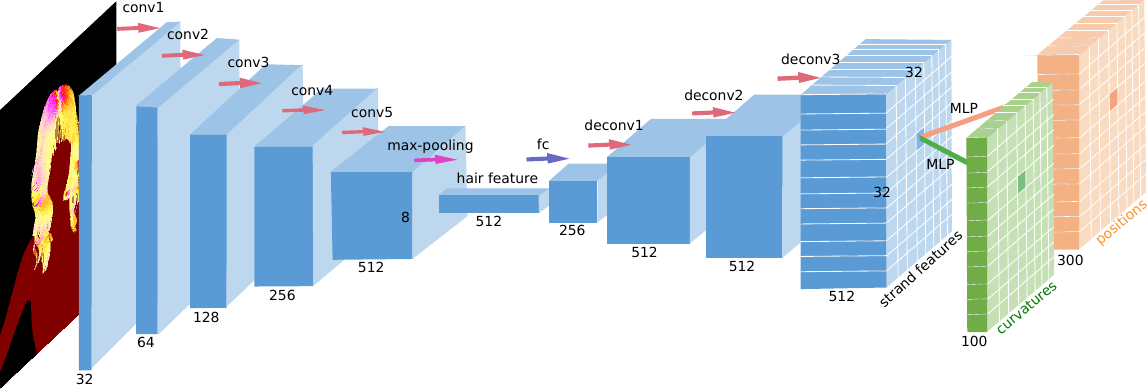}
    \caption{Network Architecture. The input orientation image is first encoded into a high-level hair feature vector, which is then decoded to $32\times32$ individual strand-features. Each strand-feature is further decoded to the final strand geometry containing both sample positions and curvatures via two multi-layer perceptron (MLP) networks.}
    \label{fg:architecture}
\end{figure}

\vspace{-20px}
\vspace{-10px}
\subsubsection{Network Architecture.}
As illustrated in Figure~\ref{fg:architecture}, our network first encodes the input image to a latent vector, followed by decoding the target hair strands from the vector.
For the encoder, we use the convolutional layers to extract the high-level features of the image. Different from the common practices that use a fully-connected layer as the last layer, we use the 2D max-pooling to spatially aggregate the partial features (a total number of $8\times8$) into a global feature vector $z$. This greatly reduces the number of network parameters.

The decoder generates the hair strands in two steps. 
The hair feature vector $z$ is first decoded into multiple strand feature vectors $\{z_i\}^{M}_{i=0}$ via deconvolutional layers, and each $z_i$ could be further decoded into the final strand geometry $\zeta$ via the same multi-layer fully connected network.
This multi-scale decoding mechanism allows us to efficiently produce denser hair models by interpolating the strand features.
According to our experiments, this achieves a more natural appearance than directly interpolating final strand geometry.

It is widely observed that generative neural networks often lose high frequency details, as the low frequency components often dominates the loss in training.
Thus, apart from the 3D position $\{\pos_{i}\}$ of each strand, our strand decoder also predicts the curvatures $\{\curv_{i}\}$ of all samples. 
With the curvature information, we can reconstruct the high frequency strand details.

\vspace{-10px}
\subsubsection{Loss Functions.}

\begin{wrapfigure}{r}{0.22\textwidth}
  \vspace*{-12.5mm}
  %\hspace*{-2.2mm}
	\begin{center}
    
   \includegraphics[width=1\textwidth]{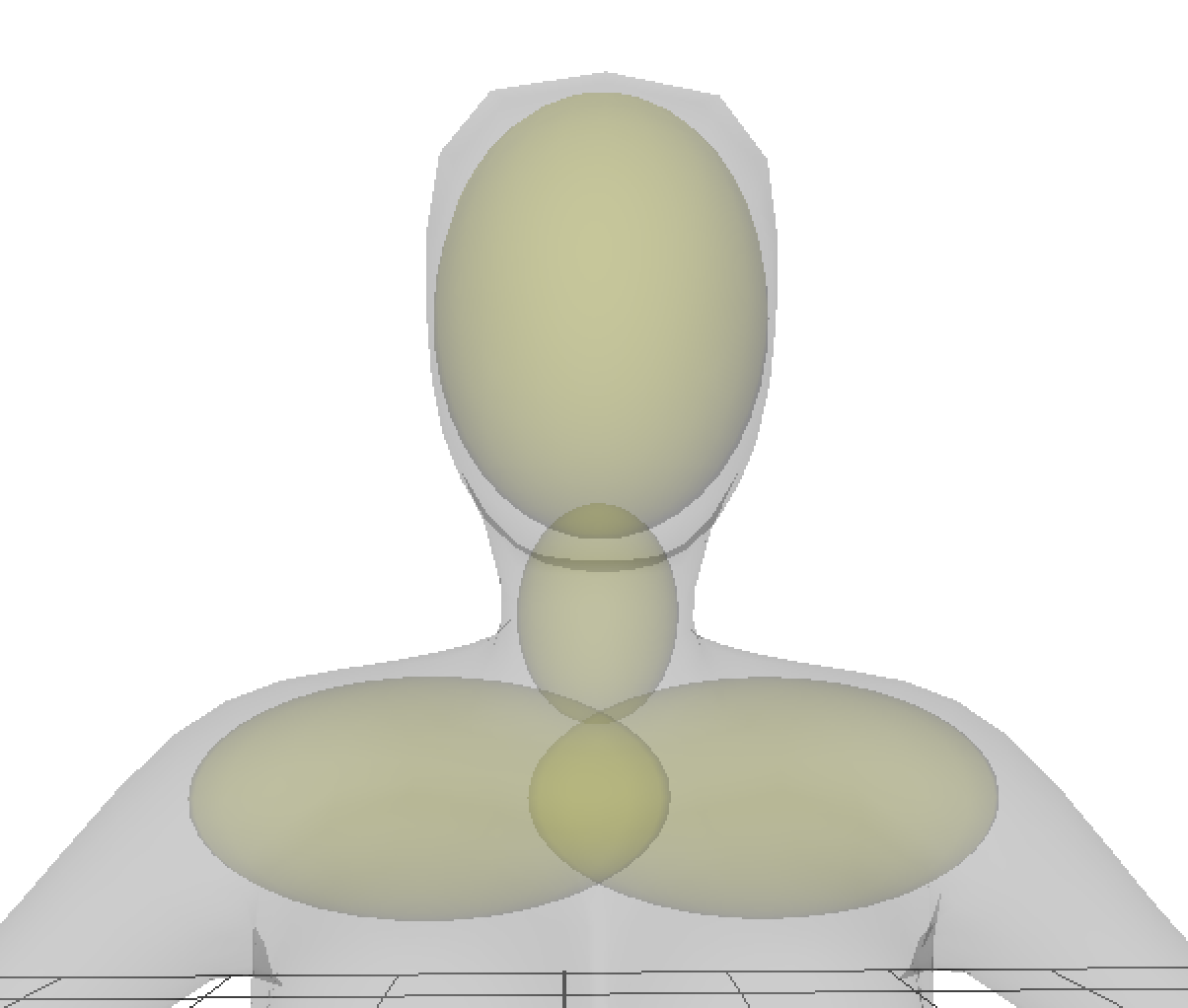}
	\end{center}
%\vspace*{-8.2mm}
\caption{Ellipsoids for Collision Test.}
\label{fg:ellipsoids}
\end{wrapfigure}

We apply three losses on our network. 
The first two losses are the $L_2$ reconstruction loss of the 3D position and the curvature of each sample. The third one is the collision loss between the output hair strand and the human body. To speed up the collision computation, we approximate the geometry of the body with four ellipsoids as shown in Figure~\ref{fg:ellipsoids}.

Given a single-view image, the shape of the visible part of the hair is more reliable than the invisible part, e.g. the inner and back hair. Thus we assign adaptive weights to the samples based on their visibility --- visible samples will have higher weights than the invisible ones. 

The final loss function is given by:
\begin{equation}
L = L_{pos} +  \lambda_{1}L_{curv} +  \lambda_{2}L_{collision}. 
\label{egn:losses}
\end{equation}
$L_{pos}$ and $L_{curv}$ are the loss of the 3D positions and the curvatures respectively, written as:
\begin{align}
\label{eq:poscurvloss}
\begin{split}
L_{pos} =  \frac{1}{NM}\sum_{i=0}^{N-1}\sum_{j=0}^{M-1}
w_{i,j}||\pos_{i,j}-\pos_{i,j}^*||_2^2\\
L_{curv} =  \frac{1}{NM}\sum_{i=0}^{N-1}\sum_{j=0}^{M-1} w_{i,j}(\curv_{i,j}-\curv_{i,j}^*)^2\\
w_{i,j} = \begin{cases}
    10.0 & \sample_{i,j} \ is \ visible \\
    0.1 & \mathrm{otherwise}
  \end{cases}
\end{split}
\end{align}
where $\pos_{i,j}^*$ and $\curv_{i,j}^*$ are the corresponding ground truth position and curvature to $\pos_{i,j}$ and $\curv_{i,j}$, and $w_{i,j}$ is the visibility weight.

\vspace{2mm}
The collision loss $L_{col}$ is written as the sum of each collision error on the four ellipsoids:
\begin{equation}
\label{eq:collision}
L_{col} =   \frac{1}{NM}\sum_{k=0}^3 C_k
\end{equation}
Each collision error is calculated as the sum of the distance of each collided point to the ellipsoid surface weighted by the length of strand that is inside the ellipsoid, written
\begin{equation}
C_{k} = \sum_{i=0}^{N-1}\sum_{j=1}^{M-1} \|\pos_{i,j} - \pos_{i, j-1}\| max(0, Dist_k)
\end{equation}
\begin{equation}
Dist_k = 1-\frac{(\pos_{i,j,0}-x_k)^2}{a_k^2}-\frac{(\pos_{i,j,1}-y_k)^2}{b_k^2}-\frac{(\pos_{i,j,2}-z_k)^2}{c_k^2}
\end{equation}
where $\|\pos_{i,j} - \pos_{i, j-1}\|$ is the $L_1$ distance between two adjacent samples on the strand. $x_k$, $y_k$, $z_k$, $a_k$, $b_k$, and $d_k$ are the model parameters of the ellipsoid.

\vspace{-10px}

\subsubsection{Training Details.}
The training parameters of Equation~\ref{egn:losses} are fixed to be $\lambda_{1}=1.0$ and $\lambda_{2}=10^{-4}$.
During training, we resize all the hair so that the hair is measured in the metric system. We use Relu for nonlinear activation, Adam \cite{kingma2014adam} for optimization, and run the training for 500 epochs using a batch size of 32 and learning rate of $10^{-4}$ multiplied by 2 after 250 epochs.

\begin{figure}[h]
    \centering
    \includegraphics[width=0.8\textwidth]{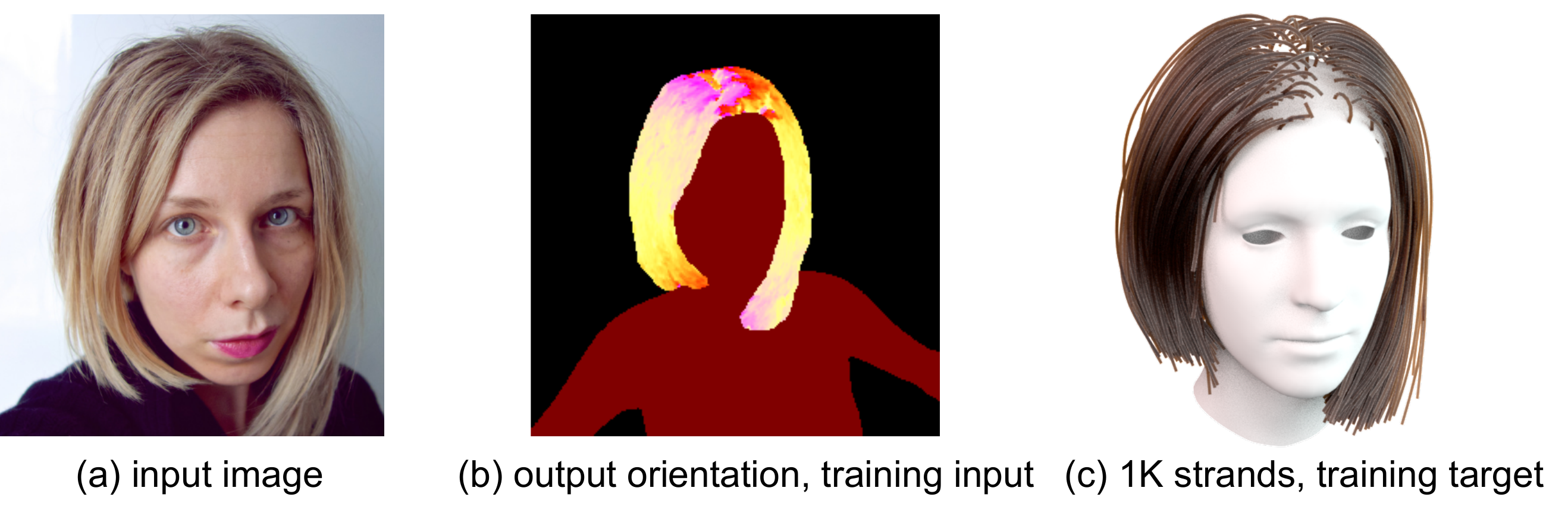}
    \caption{The orientation image (b) can be automatically generated from a real image (a), or from a synthesized hair model with 9K strands. The orientation map and a down-sampled hair model with 1K strands (c) are used to train the neural network.}
    \label{fig:curve_reconstruction}
\end{figure}

\subsection{Reconstruction}

%\vspace{-10px}
The output strands from the network may contain noise, and sometimes lose high-frequency details when the target hair is curly. Thus, we further refine the smoothness and curliness of the hair. We first smooth the hair strands by using a Gaussian filter to remove the noise.
Then, we compare the difference between the predicted curvatures and the curvatures of the output strands. If the difference is higher than a threshold, we add offsets to the strands samples. 
In particular, we first construct a local coordinate frame at each sample with one axis along the tangent of the strand, then apply an offset function along the other two axises by applying the curve generation function described in the work of Zhou et. al \cite{yu2001modeling}.
\vspace{-10px}

\begin{figure}[h!]
    \centering
    \includegraphics[width=0.8\textwidth]{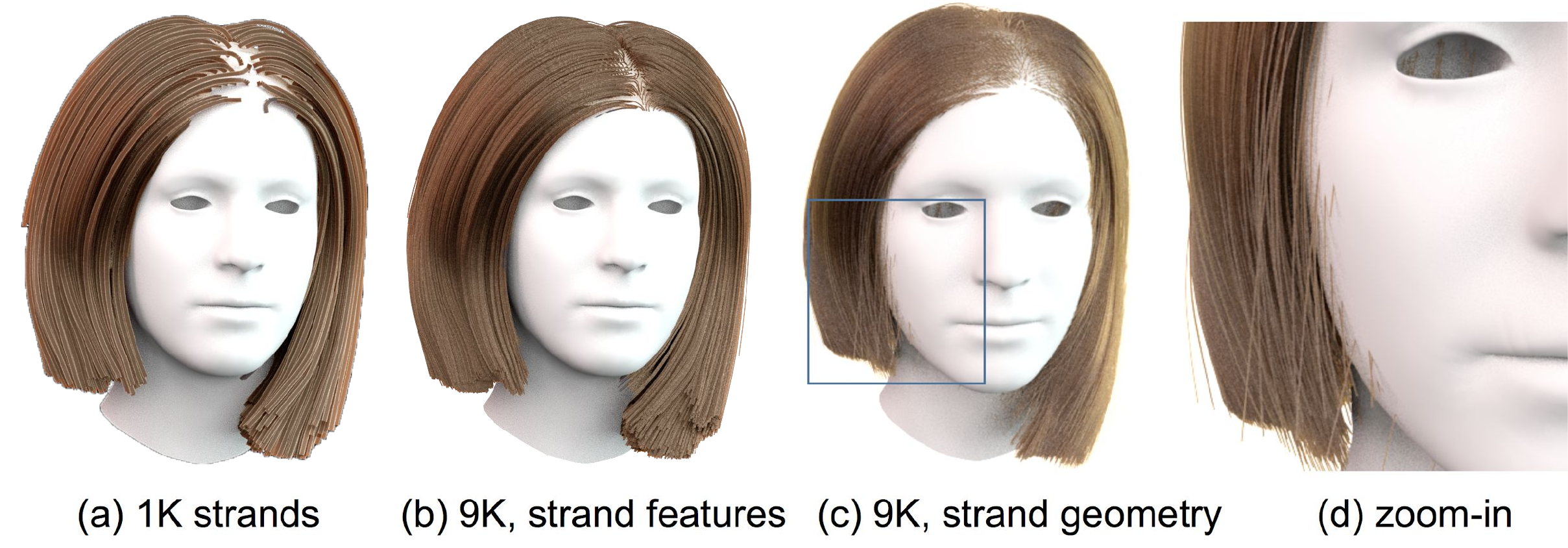}
    \vspace{-5px}
    \caption{Hair strand upsampling in the space of (b) the strand-features and (c) the final strand geometry. (d) shows the zoom-in of (c).}
    \vspace{-5px}
    \label{fig:result_upsample}
\end{figure}
\vspace{-10px}

The network only generates 1K hair strands, which is insufficient to render a high fidelity output.
To obtain higher resolution, traditional methods build a 3D direction field from the guide strands and regrows strands using the direction field from a dense set of follicles.
However, this method is time consuming and cannot be used to reconstruct an accurate hair model.
Although directly interpolating the hair strands is fast, it can also produce an unnatural appearance.
Instead, we bilinearly interpolate the intermediate strand features $z_i$ generated by our network and decode them to strands by using the perceptron network, which enables us to create hair models with arbitrary resolution.

Figure~\ref{fig:result_upsample} demonstrates that by interpolating in strand-feature space, we can generate a more plausible hair model.
In contrast, direct interpolation of the final strands could lead to artifacts like collisions. This is easy to understand, as the strand-feature could be seen as a non-linear mapping of the strand, and could fall in a more plausible space.
\vspace{-10px}

\begin{figure}[h!]
    \centering
    \includegraphics[width=0.6\textwidth]{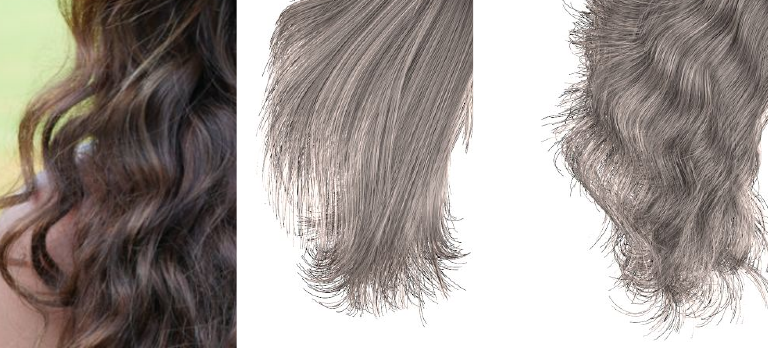}
    \caption{Reconstruction with and without using curliness.}
    \vspace{-5px}
    \label{fig:curve_reconstruction}
\end{figure}
\vspace{-15px}
Figure~\ref{fig:curve_reconstruction} demonstrates the effectiveness of adding curliness in our network. Without using the curliness as an extra constraint, the network only learns the dominant main growing direction while losing the high-frequency details.
In this paper, we demonstrate all our results at a resolution of 9K to 30K strands.

\vspace{-5px}
\section{Evaluation}

\subsection{Quantitative Results and Ablation Study}
In order to quantitatively estimate the accuracy of our method, we prepare a synthetic test set with 100 random hair models and 4 images rendered from random views for each hair model. We compute the reconstruction errors on both the visible and invisible part of the hair separately using the mean square distance between points and the collision error using equation \ref{eq:collision}.
We compare our result with Chai et al.'s method \cite{chai2016autohair}. Their method first queries for the nearest neighbor in the database and then performs a refinement process which globally deforms the hair using the 2D boundary constraints and the 2D orientation constraints based on the input image. To ensure the fairness and efficiency of the comparison, we use the same database in our training set for the nearest neighbor query of \cite{chai2016autohair} based on the visible part of the hair, and set the resolution at 1000 strands. We also compare with Hu et al.'s method \cite{hu2015single} which requires manual strokes for generating the 3D hair model. But drawing strokes for the whole test set is too laborious, so in our test, we use three synthetic strokes randomly sampled from the ground-truth model as input. In Table \ref{tb:error}, we show the error comparison with the nearest neighbor query results and the methods of both papers. We also perform an ablation test by respectively eliminating the visibility-adaptive weights, the collision loss and the curvature loss from our network. 

From the experiments, we observe that our method outperforms all the ablation methods and Chai et al.'s method. Without the visibility-adaptive weights, the reconstruction error is about the same for both the visible and invisible parts, while the reconstruction error of the visible hair decreased by around 30\% for all the networks that applies the visibility-adaptive weights. The curvature loss also helps decrease the mean square distance error of the reconstruction. The experiment also shows that using the collision loss will lead to much less error in collision. The nearest-neighbor method results have 0 collision error because the hairs in the database have no collisions.

In Table \ref{tb:time}, we compare the computation time and hard disk usage of our method and the data-driven method at the resolution of 9K strands. It can be seen that our method can be about three magnitude faster faster and only uses a small amount of storage space. The reconstruction time differs from straight hair styles and curly hair styles because for straight hair styles which have less curvature difference, we skip the process of adding curves. 

\vspace{-20px}
\begin{table}[h]
\centering
\caption{Reconstruction Error Comparison. The errors are measured in metric. The Pos Error refers to the mean square distance error between the ground-truth and the predicted hair. "-VAW" refers to eliminating the visibility-adaptive weights. "-Col" refers to eliminating the collision loss, "-Curv" refers to eliminating the curvature loss. "NN" refers to nearest neighbor query based on the visible part of the hair.}
\label{tb:error}
\begin{tabular}{|l|l|l|l|}
\hline
               & Visible Pos Error & Invisible Pos Error & Collision Error     \\ \hline
HairNet        & 0.017             & 0.027               & $2.26\times10^{-7}$ \\ \hline
HairNet - VAW  & 0.024             & 0.026               & $3.5\times10^{-7}$  \\ \hline
HairNet - Col  & 0.019             & 0.027               & $3.26\times10^{-6}$ \\ \hline
NairNet - Curv & 0.020             & 0.029               & $3.3\times10^{-7}$  \\ \hline
NN             & 0.033             & 0.041               & 0                   \\ \hline
Chai et al.\cite{chai2016autohair}         & 0.021             & 0.040               & 0                   \\ \hline
Hu et al.\cite{hu2015single}         & 0.023             & 0.028               & 0                   \\ \hline
\end{tabular}
\end{table}
\vspace{-35px}

\setlength{\tabcolsep}{0.2em}
\begin{table}[h]
\centering
\caption{Time and space complexity.}
\label{tb:time}
\begin{tabular}{|l|l|l|l|l|l|}
\hline
ours         & preprocessing & inference & reconstruction & total time & total space \\ \cline{2-6} 
             & 0.02 s       & 0.01 s        &   0.01 - 0.05 s  &  0.04 - 0.08 s   & 70 MiB      \\ \hline
Chai et al.\cite{chai2016autohair} & preprocessing & NN query & refinement & total time & total space \\ \cline{2-6} 
             & 3 s           & 10 s          & 40 s       & 53 s       & 1 TiB     \\ \hline
\end{tabular}
\end{table}

\vspace{-35px}

\subsection{Qualitative Results}

To demonstrate the generality of our method, we tested with different real portrait photographs as input, as shown in Figure~\ref{fig:result_gallery0},~\ref{fig:result_gallery1} and~\ref{fig:result_gallery2}. Our method can handle different overall shapes (e.g. short hairstyles and long hairstyles). In addition, our method can also reconstruct different levels curliness within hairstyles (e.g. straight, wavy, and very curly) efficiently, since we learn the curliness as curvatures in the network and use it to synthesize our final strands.

\begin{figure}[h!]
    \centering
    \includegraphics[width=1.0\textwidth]{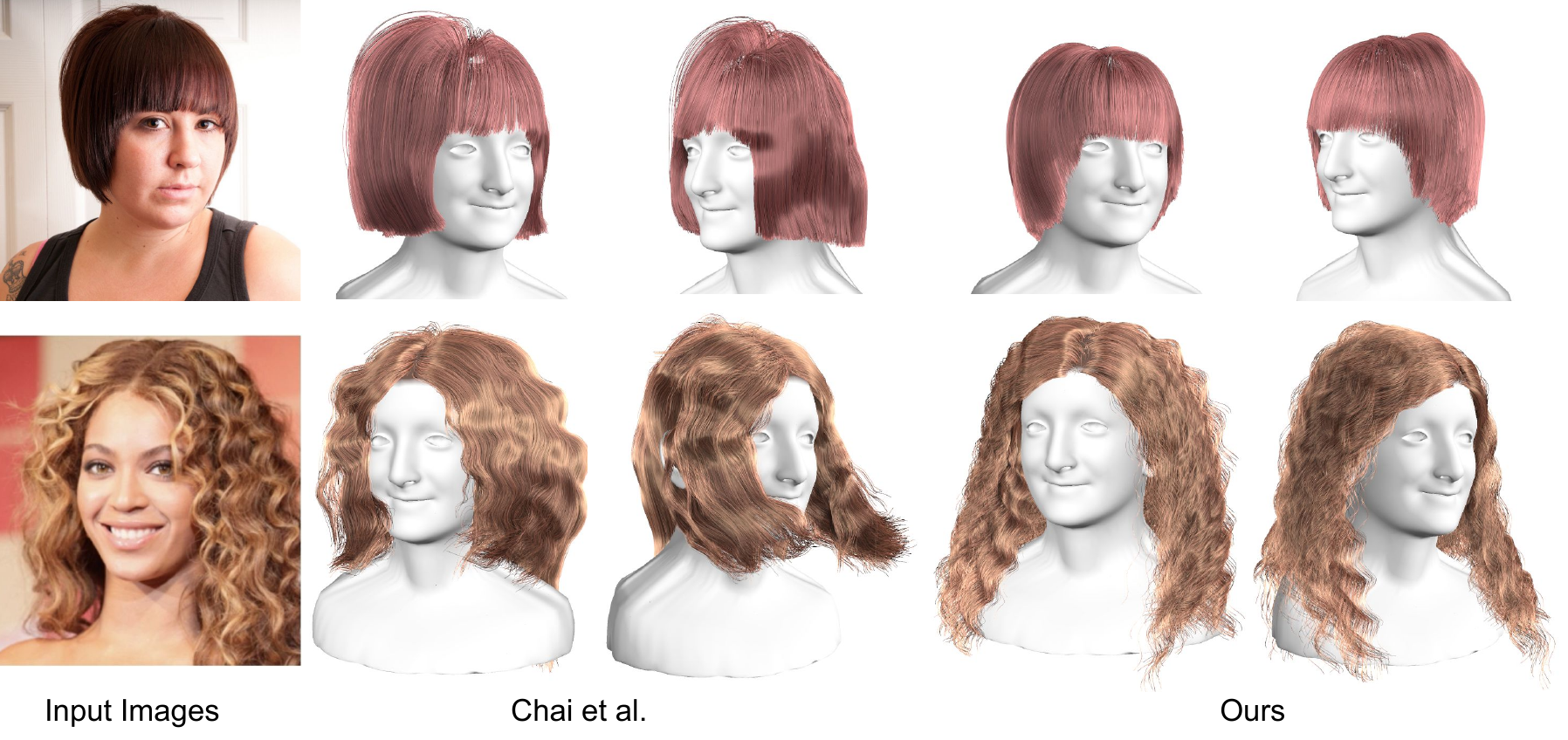}
    \vspace{-5px}
    \caption{Comparison with Autohair in different views~\cite{chai2016autohair}.}
    \vspace{-5px}
    \label{fig:visual_comparison_views}
\end{figure}

\begin{figure}[h!]
    \centering
    \includegraphics[width=1.0\textwidth]{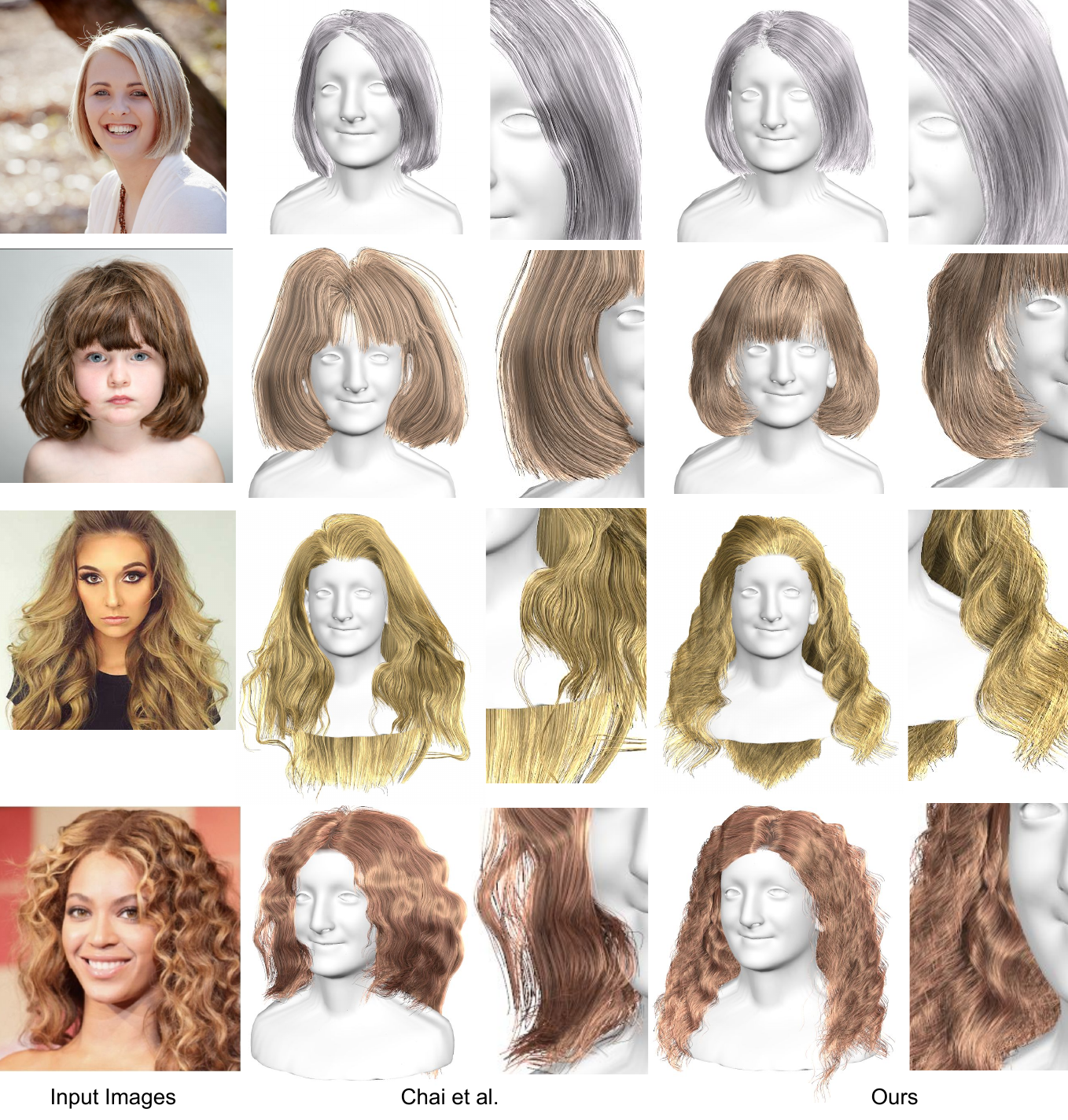}
    \vspace{-5px}
    \caption{Comparison with Autohair for local details.~\cite{chai2016autohair}.}
    \vspace{-5px}
    \label{fig:visual_comparison_details}
\end{figure}

In Figure~\ref{fig:visual_comparison_details} and Figure~\ref{fig:visual_comparison_views}, we compare our results of single-view hair reconstruction with autohair \cite{chai2016autohair}. We found that both methods can make rational inference of the overall hair geometry in terms of length and shape, but the hair from our method can preserve better local details and looks more natural, especially for curly hairs. This is because Chai et al.'s method depends on the accuracy and precision of the orientation field generated from the input image, but the orientation field generated from many curly hair images is noisy and the wisps overlap with each other. 
In addition, they use helix fitting to infer the depth of the hair, but it may fail for very curly hairs, as shown in the second row of Figure~\ref{fig:visual_comparison_views}. 
Moreover, Chai et al.'s method can only refine the visible part of the hair, so the reconstructed hair may look unnatural from views other than the view of the input image, while the hair reconstructed with our method looks comparatively more coherent from additional views.

\vspace{-15px}

\begin{figure}[h!]
    \centering
    \includegraphics[width=1.0\textwidth]{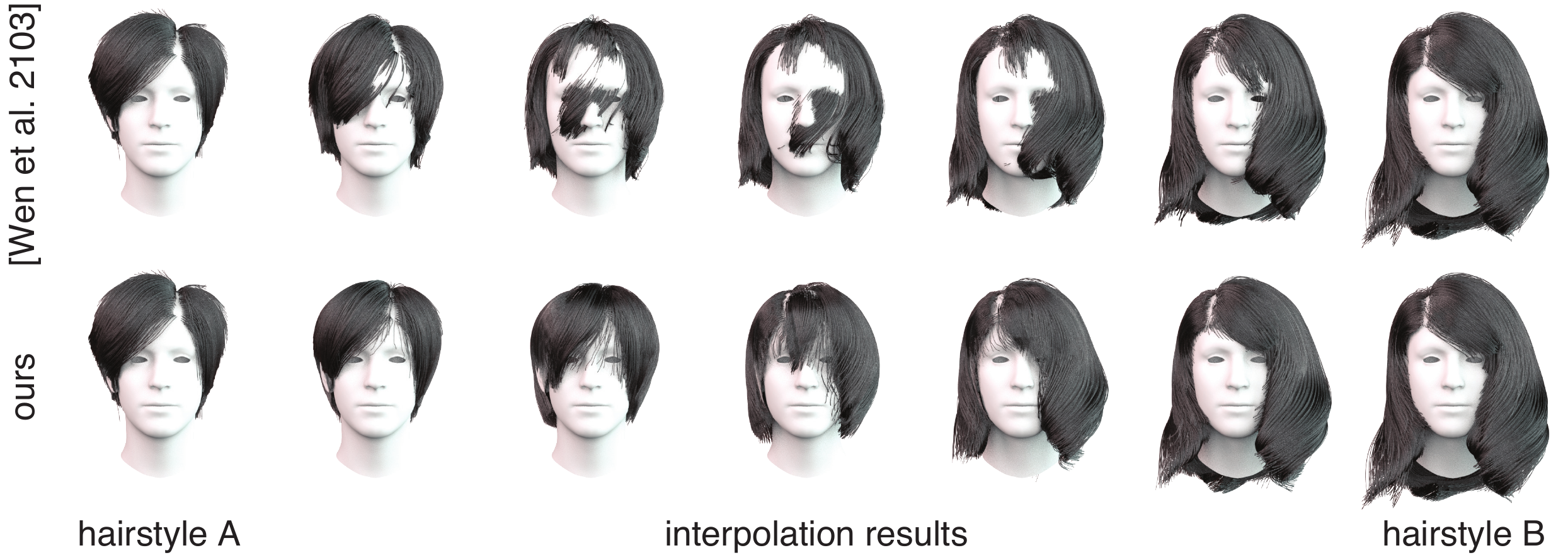}
    \vspace{-5px}
    \caption{Interpolation comparison.}
    \vspace{-5px}
    \label{fig:interpolation_gallery}
\end{figure}
\vspace{-30px}

\vspace{-15px}
Figure~\ref{fig:result_dense_interpolation} show the interpolation results of our method. The interpolation is performed between four different hair styles and the result shows that our method can smoothly interpolate hair between curly or straight and short or long hairs. We also compare interpolation with Weng et al.'s method \cite{Weng:2013:HIP}. In Figure~\ref{fig:interpolation_gallery}, Weng et al.'s method produces a lot of artifacts while our method generates more natural and smooth results. The interpolation results  indicate the effectiveness of our latent hair representation.

\begin{figure}[h]
    \centering
    \includegraphics[width=1.0\textwidth]{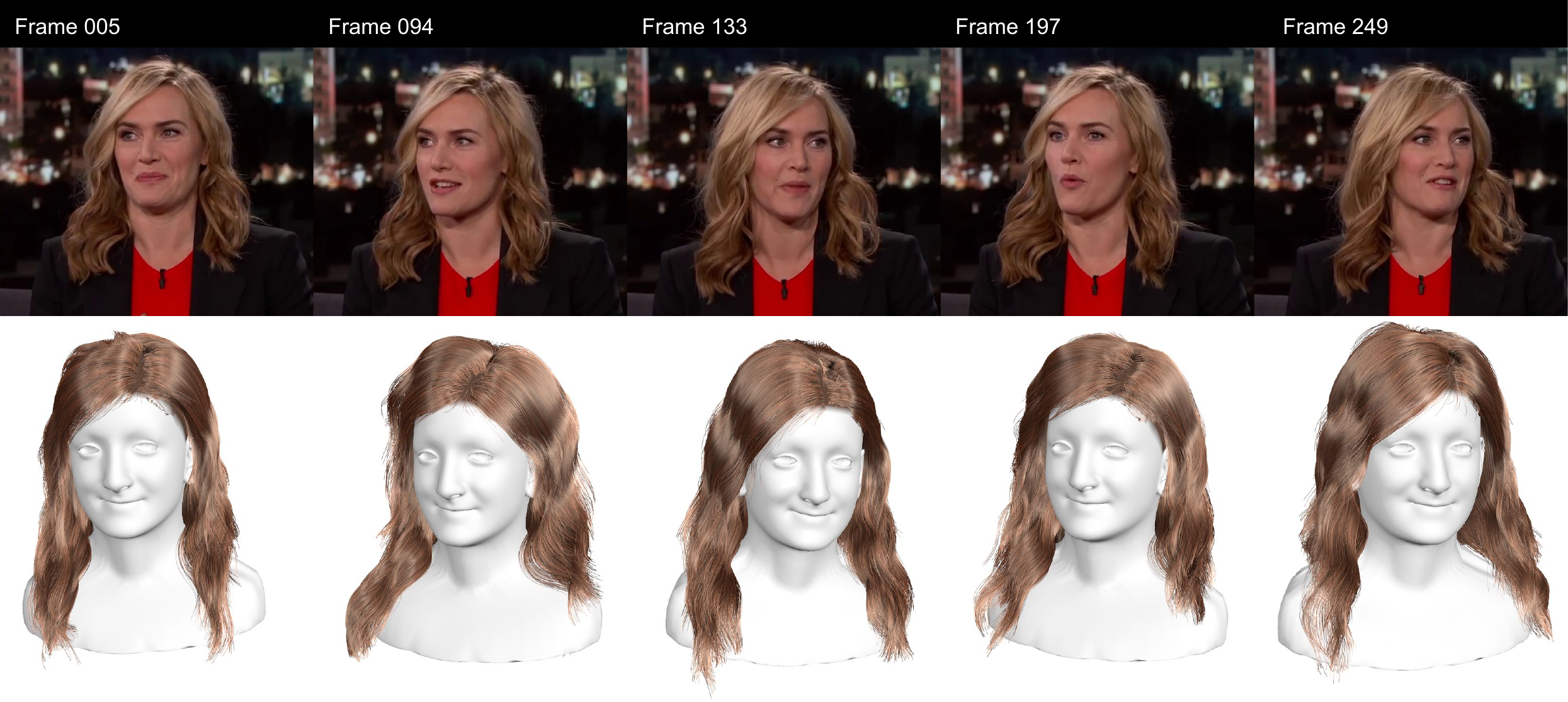}
    \vspace{-5px}
    \caption{Hair tracking and reconstruction on video.}
    \vspace{-5px}
    \label{fig:result_video}
\end{figure}

We also show video tracking results (see Figure~\ref{fig:result_video} and supplemental video).
It shows that our output may fail to achieve sufficient temporal coherence.
\vspace{-10px}

\vspace{-20px}

\vspace{-10px}
\section{Conclusion}
\vspace{-5px}
We have demonstrated the first deep convolutional neural network capable of performing real-time hair generation from a single-view image.
By training an end-to-end network to directly generate the final hair strands, our method can capture more hair details and achieve higher accuracy than current state-of-the-art.
The intermediate 2D orientation field as our network input provides flexibility, which enables our network to be used for various types of hair representations, such as images, sketches and scans given proper preprocessing.
By adopting a multi-scale decoding mechanism, our network could generate hairstyles of arbitrary resolution while maintaining a natural appearance.  
Thanks to the encoder-decoder architecture, our network provides a continuous hair representation, from which plausible hairstyles could be smoothly sampled and interpolated.

\vspace{-10px}
\section{Limitations and Future Work}
\begin{figure}[h!]
    \centering
    \includegraphics[width=1.0\textwidth]{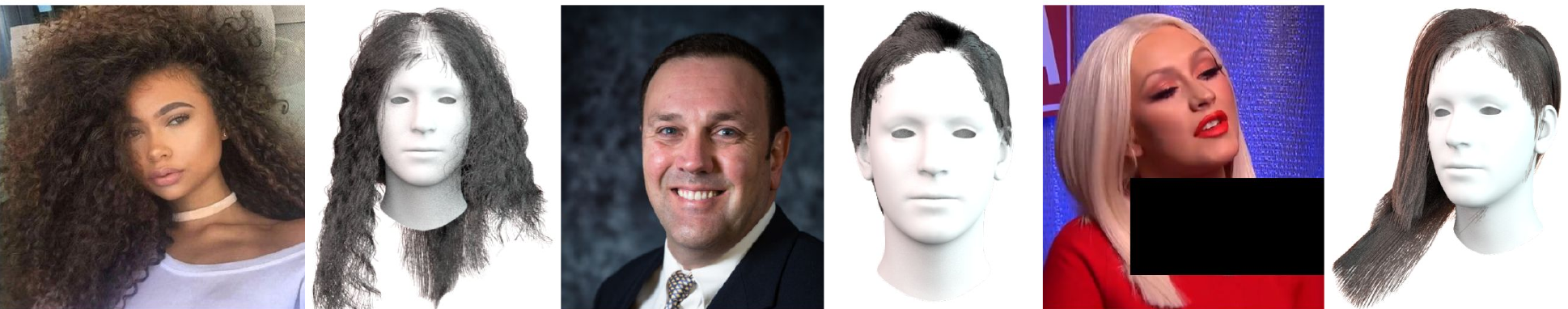}
    \caption{Failure Cases.}
    \label{fig:failure_case}
\end{figure}

\vspace{-5px}
We found that our approach fails to generate exotic hairstyles like kinky, afro or buzz cuts as shown in Figure~\ref{fig:failure_case}. We think the main reason is that we do not have such  hairstyles in our training database.
Building a large hair dataset that covers more variations could mitigate this problem.
Our method would also fail when the hair is partially occluded.
Thus we plan to enhance our training in the future by adding random occlusions.
In addition, we use face detection to estimate the pose of the torso in this paper, but it can be replaced by using deep learning to segment the head and body. % since our network actually takes the 2D mask of the torso as input. 
Currently, the generated hair model is insufficiently temporally coherent for video frames. 
Integrating temporal smoothness as a constraint for training is also an interesting future direction.
Although our network provides a more compact representation for the hair, 
there is no semantic meaning of such latent representation.
It would be interesting to concatenate explicit labels (e.g. color) to the latent variable for controlled training.
\vspace{-10px}

\vspace{-25px}

\vspace{-10px}

\vspace{-20px}
%\clearpage

\bibliographystyle{splncs}
\bibliography{egbib}

\appendix

\section{Hair Interpolation}

\begin{figure}[h!]
    \centering
    \includegraphics[width=1.0\textwidth]{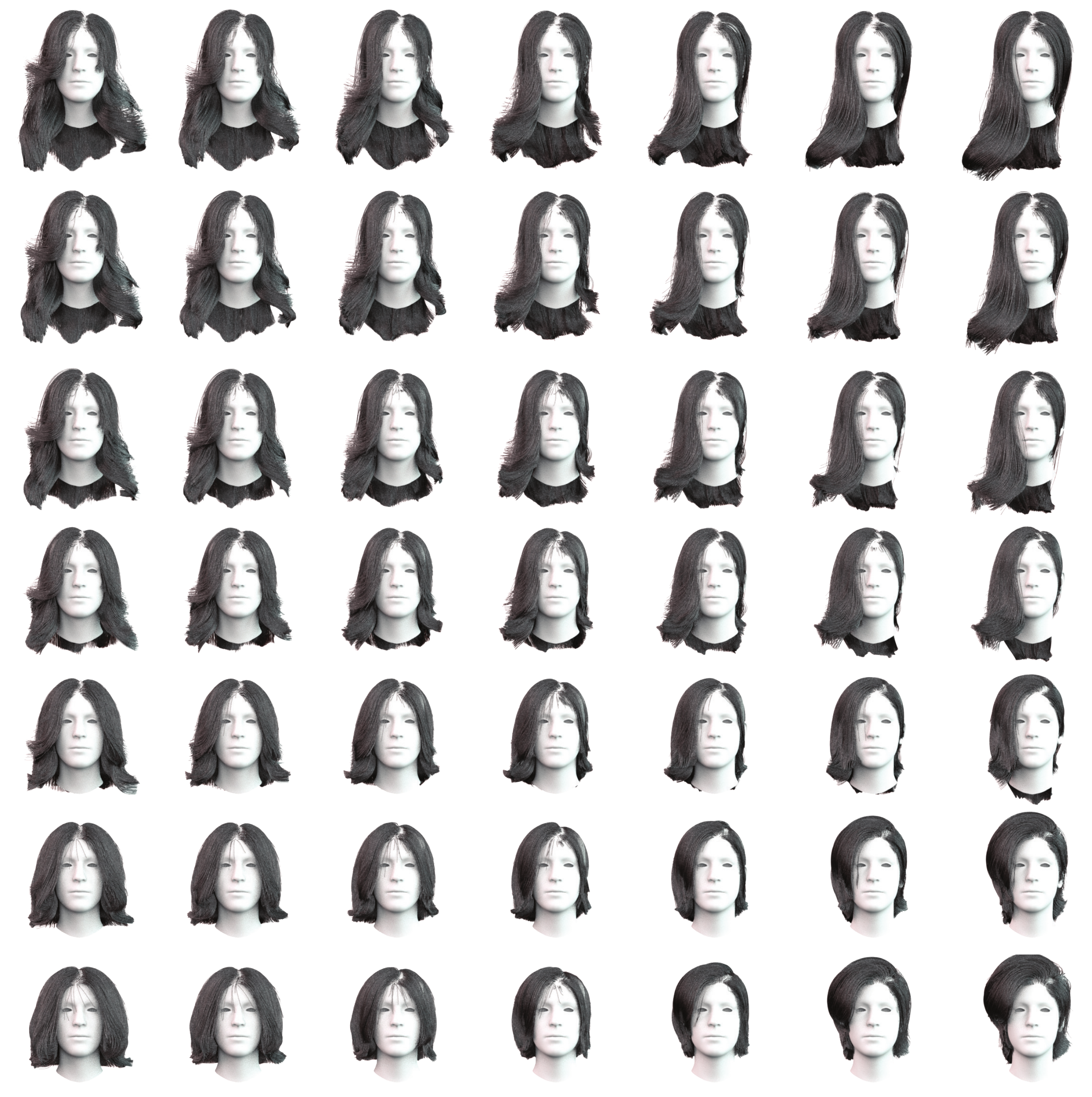}
    \vspace{-5px}
    \caption{Hair interpolation in the latent space of our model. Given the four corner hairs with various shapes and lengths, more new hairs can be generated by interpolating their latent vectors from the encoder.  Each output's position in the above layout reflects its relative weights for the corresponding input models.}
    \vspace{-5px}
    \label{fig:result_dense_interpolation}
\end{figure}
\clearpage

\section{Detailed Network Architecture}
\vspace{-20px}
\begin{table}[]
%\centering
\caption{Details of our network architecture. "in\_ch" means input channel size. "out\_ch" means output channel size.}
\label{my-label}
\begin{tabular}{l}
\hline
Encoder (input: 2 $\times$ 256 $\times$ 256, output: 512)                                         \\ \hline
Conv2d(in\_ch=3, out\_ch=32, kernel\_size=8, stride=2, padding=3)    ReLU                                                                                         \\
Conv2d(in\_ch=32, out\_ch=64, kernel\_size=8, stride=2, padding=3)  ReLU                                                                                         \\
Conv2d(in\_ch=64,out\_ch=128, kernel\_size=6, stride=2, padding=2)  ReLU                                                                                         \\
Conv2d(in\_ch=128, out\_ch=256, kernel\_size=4, stride=2, padding=1) ReLU                                                                                         \\
Conv2d(in\_ch=256, out\_ch=256, kernel\_size=3, stride=1, padding=1) ReLU                                                                                         \\
Conv2d(in\_ch=256, out\_ch=512, kernel\_size=4, stride=2, padding=1) ReLU                                                                                         \\
Conv2d(in\_ch=512, out\_ch=512, kernel\_size=3, stride=1, padding=1) ReLU                                                                                         \\
MaxPool2d(kernel\_size=8) Tanh                                                                                         \\ \hline
Decoder(input: 512, output: 512 $\times$ 32 $\times$ 32)                                         \\ \hline
Linear(512, 1024)  ReLU                                                                                         \\
Linear(1024, 4096) ReLU                                                                       \\
Bilinear Upsample(scale\_factor=2)                                                           \\
Conv2d(in\_ch=256, out\_ch=512, kernel\_size=3, stride=1, padding=1) ReLU                                                                                         \\
Bilinear Upsample(scale\_factor=2)                                                           \\
Conv2d(in\_ch=512, out\_ch=512, kernel\_size=3, stride=1, padding=1) ReLU                                                                                         \\
Bilinear Upsample(scale\_factor=2)                                                           \\
Conv2d(in\_ch=512, out\_ch=512, kernel\_size=3, stride=1, padding=1) ReLU                                                                                         \\ \hline
Strand Curvature Decoder(input: 512 $\times$ 32 $\times$ 32, output 100 $\times$ 32 $\times$ 32)                                                              \\ \hline
Conv2d(in\_ch=512, out\_ch=512, kernel\_size=1, stride=1, padding=0) ReLU  \\
Conv2d(in\_ch=512, out\_ch=512, kernel\_size=1, stride=1, padding=0) Tanh\\
Conv2d(in\_ch=512, out\_ch=100, kernel\_size=1, stride=1, padding=0)\\
\hline
Strand Position Decoder(input: 512 $\times$ 32 $\times$ 32, output 300 $\times$ 32 $\times$ 32)                                                              \\ \hline
Conv2d(in\_ch=512, out\_ch=512, kernel\_size=1, stride=1, padding=0) ReLU  \\
Conv2d(in\_ch=512, out\_ch=512, kernel\_size=1, stride=1, padding=0) Tanh\\
Conv2d(in\_ch=512, out\_ch=300, kernel\_size=1, stride=1, padding=0)\\
                                                                                            
\end{tabular}
\end{table}
\vspace{-30px}
\section{Collision}
\begin{figure}[h!]
    \centering
    \includegraphics[width=0.8\textwidth]{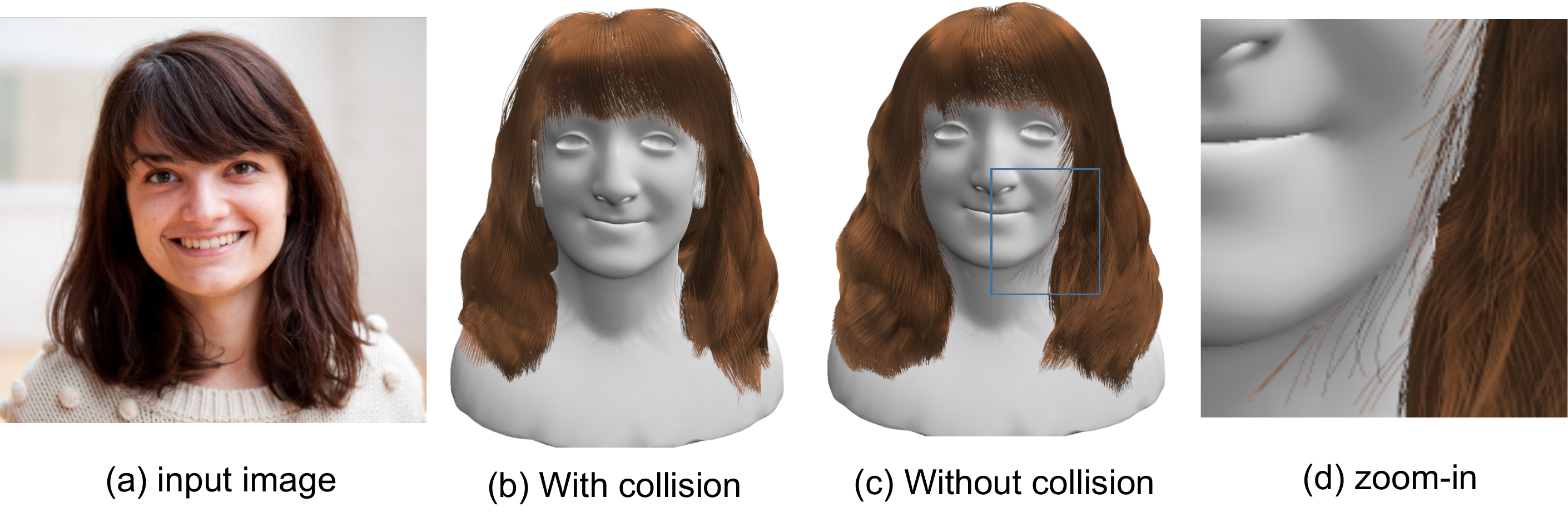}
    \caption{Without the collision loss in training (c,d), our method could produce less plausible results.}
    \vspace{-5px}
    \label{fig:collision}
\end{figure}

\vspace{-10px}
In Figure~\ref{fig:collision}, we show the results with and without using collision loss in our training process.
Even when a simplified body model is used for collision detection, it could push the generated hair towards a more plausible space.
\vspace{-20px}

%\vspace{-30px}
\section{Results Gallery}

\begin{figure}[h!]
    \centering
    \includegraphics[width=1.0\textwidth]{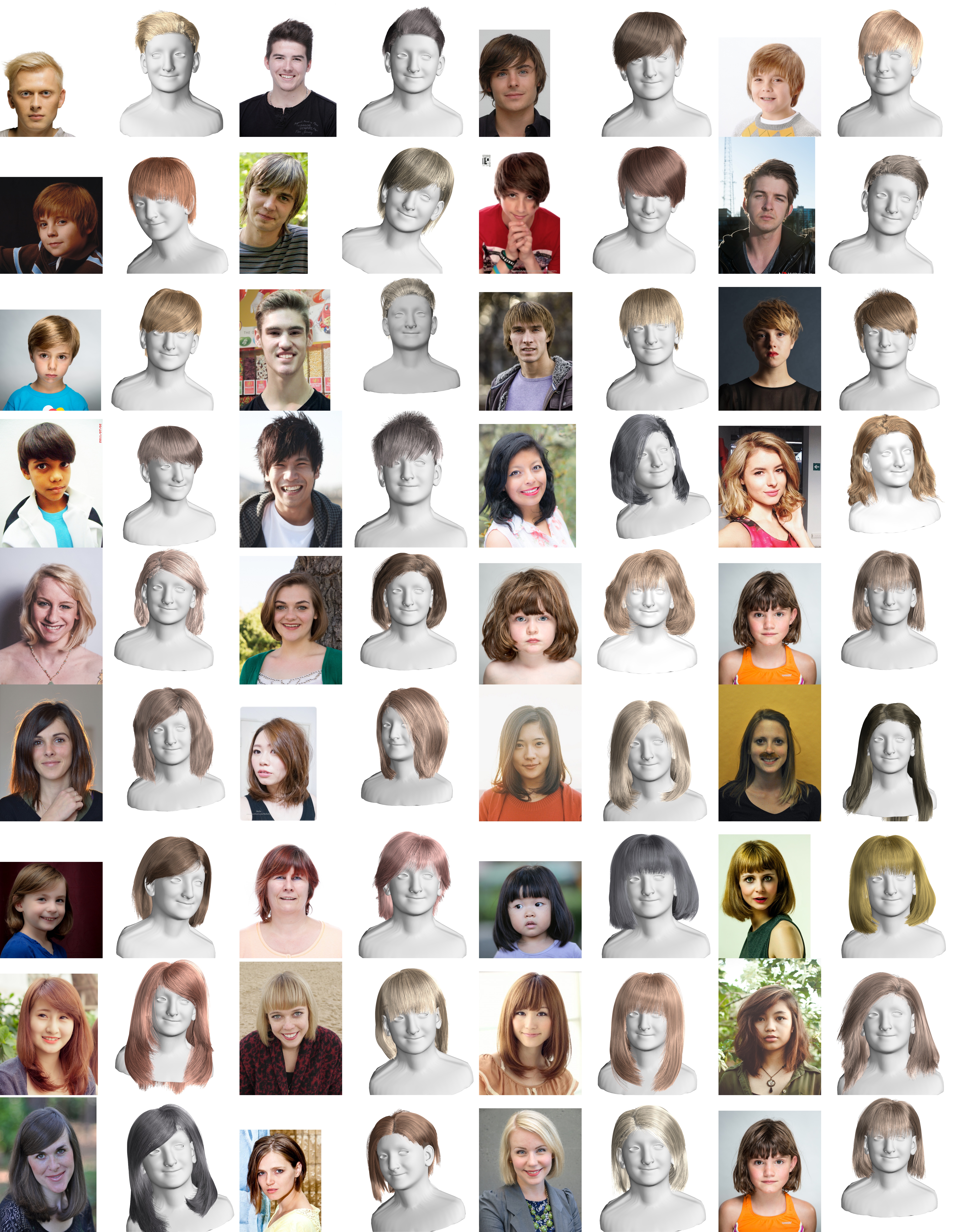}
    \caption{Our single-view reconstruction results for various hairstyles.}
    \label{fig:result_gallery0}
\end{figure}

\begin{figure}[h!]
    \centering
    \includegraphics[width=1.0\textwidth]{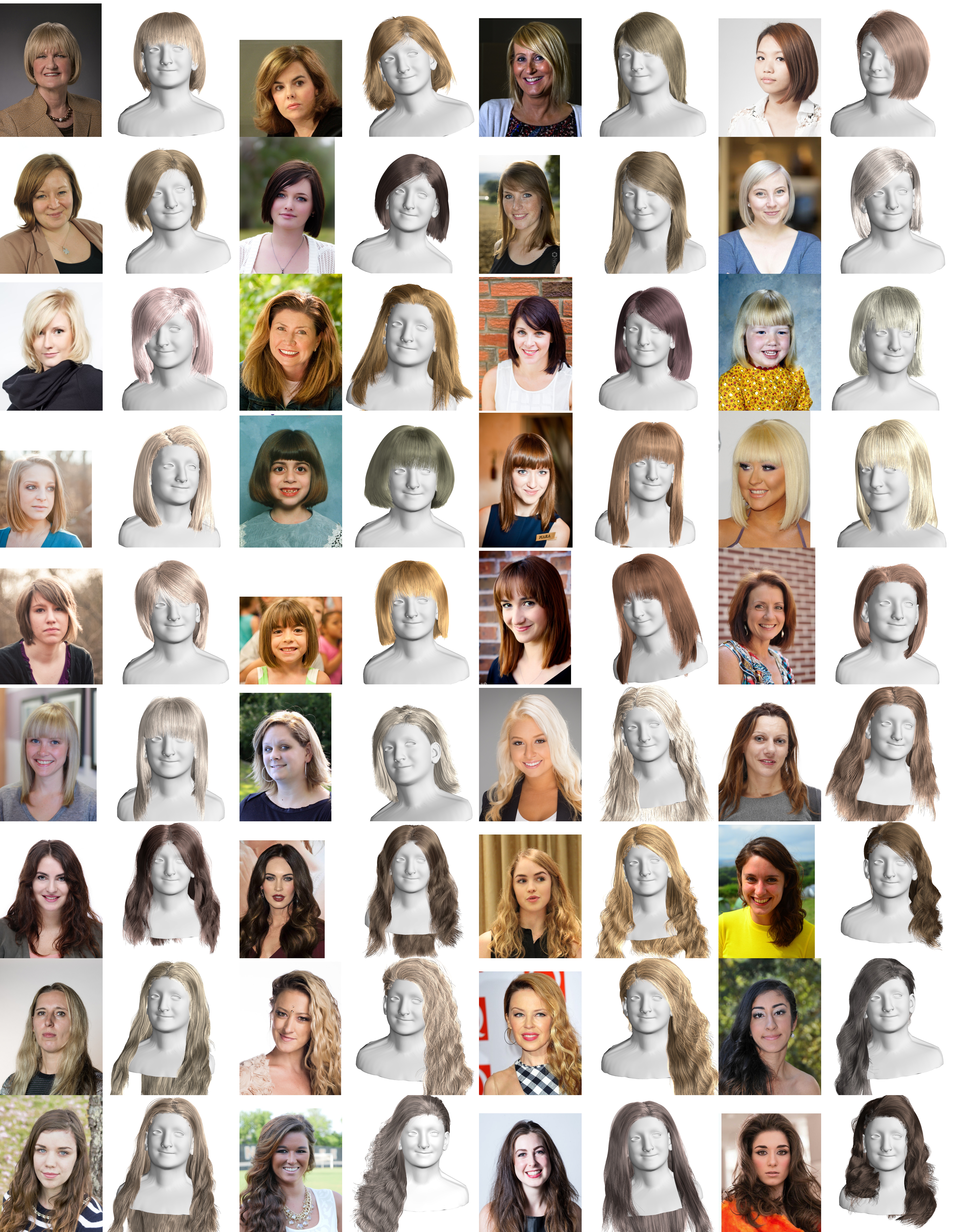}
    \caption{Our single-view reconstruction results for various hairstyles.}
    \label{fig:result_gallery1}
\end{figure}

\begin{figure}[h!]
    \centering
    \includegraphics[width=1.0\textwidth]{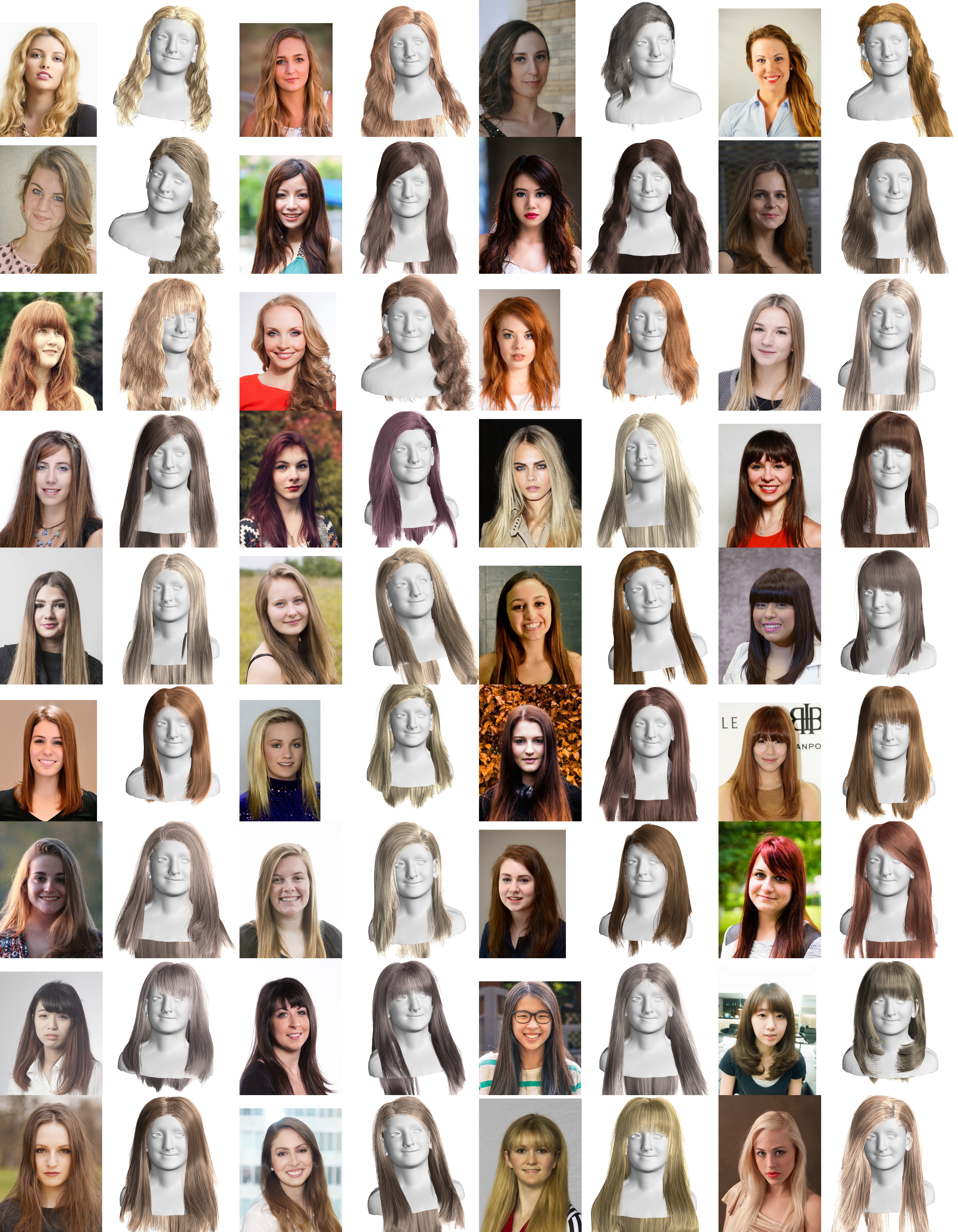}
    \caption{Our single-view reconstruction results for various hairstyles.}
    \label{fig:result_gallery2}
\end{figure}

\begin{figure}[t] 
    \includegraphics[width=1.0\textwidth]{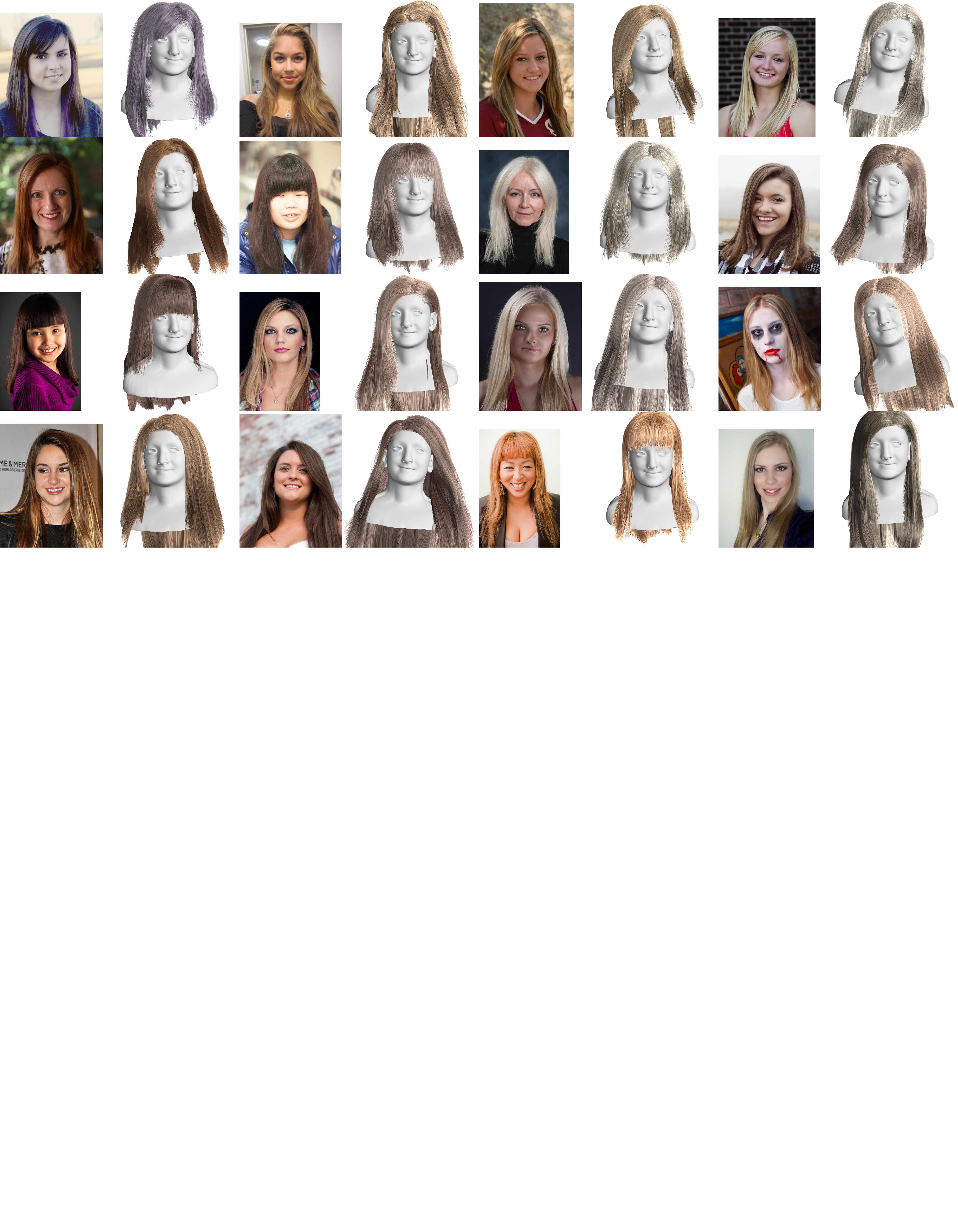}
    \vspace{-250px}
    \caption{Our single-view reconstruction results for various hairstyles.}
    \label{fig:result_gallery3}
\end{figure}

\end{document}